# Mass and the creation of spatial volume

C. L. Herzenberg

**Abstract**
The distortion of space by the presence of mass in general relativity appears to be capable of increasing three dimensional spatial volume. We examine excess volume effects associated with an isolated mass described by the Schwarzschild solution to the field equations. Fractional differential excess spatial volume in the vicinity of a mass is shown to be a direct measure of gravitational potential which can be easily visualized. The total amount of excess spatial volume associated with an isolated mass is evaluated and shown to be appreciable. Summing over the excess spatial volume contributions from individual masses present throughout the universe leads to an overall excess spatial volume comparable in magnitude to the volume of the observable universe. Interpretations of these results are discussed, including the possibility that this excess volume might contribute to clarifying the issue of missing matter in the contemporary universe. Alternatively, it is proposed speculatively that mass, rather than simply distorting preexisting space, might actually create it.

**Introduction: Where this is going**

That the presence of mass distorts the space-time around it is a well-known feature of general relativity.[1] This geometrical distortion of the surrounding space-time by the presence of mass causes the phenomenon of gravitation.

The space-time distortion produced by mass in general relativity can involve modifications in distances or lengths along some dimensions. As a result, under appropriate conditions, the mass-imposed distortion of space-time might be expected to lead to changes in the volume of three dimensional space.

The excess in spatial volume associated with the presence of objects having mass can be evaluated. Using the Schwarzschild solution to the field equations of general relativity, we will estimate the excess spatial volume originating from the distortion of space around the mass of a spherically symmetrical object. Our interest is largely in the weak field limit that would correspond to Newtonian gravitation, but we wish to explore this aspect of the concept of spatial distortion as an interpretive aid.

We estimate the differential excess volume as a function of radial distance from a mass, and find that it provides a visualizable measure of gravitational potential. We then integrate the differential excess volume to obtain the total excess volume associated with an individual mass.

We use this evaluation of the excess volume originating from an isolated individual mass as a basis for estimating the total contribution of excess volume from an assembly of objects. We then estimate the excess volume associated with all of the objects in the observable universe. When a comparison is made of this excess volume contributed by the mass of objects constituting the universe to the total volume of the observable universe, these two quantities turn out to be surprisingly similar.

Accordingly, it appears that including the contribution of mass-induced volume could lead to a universe with considerably greater spatial volume than would otherwise be anticipated. Meantime, astrophysics is at present faced with an at least partially unresolved issue, a deficit of ordinary baryonic matter, such that about half of what is expected to be seen in nearby galaxies appears to be missing. Contemporary stars, galaxies, and gas that are observed seem to account for less than half of the baryonic matter expected on the basis of modeling based on the matter visible in the early universe, so that nearly half of the known universe appears to be missing.[2,3,4] Alternatively, could dilution by an appreciable amount of excess volume contribute to the explanation of this mystery of missing matter in the contemporary universe?

The rather remarkable coincidence that the mass-induced excess spatial volume corresponds so closely to the volume of the visible universe also suggests an alternative idea that the spatial volume of the universe might actually originate from the summed result of the increments of space associated with all of the individual objects in the universe. Thus, we might speculate that the volume of space constituting our visible universe might be created by the masses present in it; that is, that masses don't just simply distort a preexistent space, rather, that they might be creating all of space itself.

**Space-time metrics: The Schwarzschild solution and the distortion of space-time around a spherically symmetric mass**

In flat three-dimensional space, the metric $ds^2$ specifies the infinitesimal distance between two points. In ordinary rectilinear Cartesian coordinates, the metric is given by $ds^2 = dx^2 + dy^2 + dz^2$, while in spherical polar coordinates, it is given by $ds^2 = dr^2 + r^2(d\theta^2 + \sin^2\theta\, d\varphi^2)$. These are the metrics that characterize ordinary empty Euclidean space.[5]

In order to describe a curved space, a curvature coefficient can be introduced as a multiplicative coefficient of the $dr^2$ term. In order to obtain a space-time metric that is compatible with special relativity, an additional term $-c^2dt^2$ must be added to account for space-time distance associated with infinitesimal temporal separations.[1] In order to obtain a space-time metric in a curved space, a coefficient describing the curvature also needs to be introduced as a multiplicative coefficient to the $-c^2dt^2$ term.[6,7]

The Schwarzschild solution is the unique solution to the field equations of general relativity in vacuum with a spherically symmetric matter distribution; it is an exact solution that describes the space-time outside a spherically symmetric, non-rotating mass.[8] We can work with this solution to the field equations so as to obtain some estimates of the relevant characteristics of the distortion of space and time in the vicinity of a mass.

The Schwarzschild solution is characterized by the metric:[1,5,6,8]

$$ds^2 = -c^2 d\tau^2 = (1 - r_s/r)^{-1} dr^2 + r^2(d\theta^2 + \sin^2\theta \, d\varphi^2) - (1 - r_s/r) c^2 dt^2 \qquad (1)$$

Here r is a measure of distance from the central mass, t is a measure of time, θ and φ are the usual angular polar coordinates. The quantity $r_s = 2GM/c^2$ is the Schwarzschild radius, a radial distance characterizing the mass M, with G designating the gravitational constant and c designating the speed of light. It can be seen that when M = 0, the Schwarzschild radius $r_s$ = 0, and the preceding equation Eqn. (1) reduces simply to an expression of the metric for special-relativistic space-time expressed in spherical polar coordinates.[6]

Our concern in this paper is with the effects of mass on the volume of our common shared three dimensional space. Consequently, we limit our attention to external spatial regions where the Schwarzschild solution is valid. We will therefore disregard behavior interior to the mass distribution of the object and also disregard behavior interior to the Schwarzschild radius of the object for present purposes.

The Schwarzschild metric can be interpreted as telling us that, in the presence of a mass and its gravitational field, time contracts and radial distance expands. Since radial distance expands in the presence of a mass, we can anticipate that the volume of three dimensional space may also be expected to increase in the presence of a mass.

**Volumetric effects in curved space: The Schwarzschild solution**

We will now attempt to develop an estimate of the extent to which the three-dimensional volume of the distorted space around a mass located at the origin of coordinates would differ from the corresponding volume of empty space.

The differential volume element in ordinary three dimensional empty Euclidean space is $dV_o$ = dx dy dz, or, in spherical coordinates,

$$dV_o = dr \, (r \, d\theta) \, (r \sin\theta \, d\varphi) \qquad (2)$$

Accordingly, if we are dealing with ordinary Euclidean three dimensional space in the absence of mass and we have a spherically symmetric situation without angular dependence, the radial differential volume element for a spherical shell of radius r, area $4\pi r^2$ and thickness dr surrounding a point at the origin of coordinates would be given by:

$$dV_o = 4\pi r^2 dr \qquad (3)$$

This expression for $dV_o$ is the measure of the differential volume associated with an increment of radial distance dr from the origin of coordinates, in empty space.

If we integrate this quantity from the origin of coordinates out to a radius R, this will lead to the familiar result that, in empty space, the total spatial volume out to a radius R is simply equal to the volume of a sphere of radius R, namely $V_o = (4\pi/3)R^3$.

We will next examine an analogous quantity for the case in which a mass is present at the origin of coordinates, distorting the space around it. In analogy with the case of undistorted space expressed in Eqn. (3), we will introduce a differential volume element $dV_m$ as a measure of the incremental volume associated with an increment of radial distance away from the mass present at the origin of coordinates:

$$dV_m = 4\pi r^2 ds \qquad (4)$$

Next, we insert into Eqn. (4) the expression for ds that can be obtained from Eqn. (1) for the case that no angular or temporal increments are involved, that is, from the equation $ds^2 = (1 - r_s/r)^{-1} dr^2$. Thus we find for what we can designate as the effective radial differential volume of space in the presence of mass:

$$dV_m = 4\pi r^2 (1 - r_s/r)^{-1/2} dr \qquad (5)$$

For use here we have developed this expression for the differential spatial volume element in the presence of mass by analogy rather than rigorously. More formal treatment and derivation of the more general volume element in space-time and the corresponding spatial volume element in the case of the Schwarzschild solution as given in Eqn. (5) can be found in the literature.[5,6,7,9,10,11]

It can be seen by comparing Eqn. (5) with Eqn. (3) that the differential volume element in the presence of mass, $dV_m$, is always larger than the differential volume element in empty space, $dV_o$, for all values of r exceeding the Schwarzschild radius. Furthermore, Eqn. (5) shows that the magnitude of $dV_m$ approaches the magnitude of $dV_o$ for large values of r, at distances far away from the centrally located mass.

Thus, we see that the Schwarzschild solution does appear to exhibit excess three dimensional spatial volume associated with the presence of mass. (This takes place even though no excess four dimensional space-time volume would seem to be predicted for the Schwarzschild solution in general relativity, since in forming the four dimensional volume element the coefficient of dt would amount to the inverse of the coefficient of dr.)[7]

For sufficiently large values of r compared to the Schwarzschild radius, the last equation, Eqn. (5), is approximated by:

$$dV_m \approx 4\pi r^2 (1 + r_s/2r + \ldots) dr \qquad (6)$$

What we are mainly interested in is the difference between the effective spatial volume in the presence of a mass and the corresponding spatial volume in the absence of the mass, which would provide a measure of the excess spatial volume introduced by the presence of the mass. Combining Eqn. (5) and Eqn. (3), we can obtain:

$$d(V_m - V_o) = dV_m - dV_o = 4\pi r^2 [(1 - r_s/r)^{-1/2} - 1] \, dr \qquad (7)$$

This differential volume difference can be approximated for sufficiently large values of r by:

$$d(V_m - V_o) \approx 4\pi r^2 [(1 + r_s/2r + \ldots) - 1] \, dr \approx (2\pi r_s) r \, dr \qquad (8)$$

When we examine the differential of the volume difference in a concentric spherical shell as given in Eqn. (8), we see that the contribution to the three dimensional spatial volume will be greater if a centrally located mass is present than if it is absent.

**Differential excess volume and gravitational potential**

Next, let's evaluate the ratio of the differential excess volume in the presence of a mass to the differential volume in the absence of a mass. This fraction will give us a measure of the differential spatial volume caused by the presence of mass. Combining Eqn. (8) for the differential excess volume with Eqn. (3) for the differential volume in the absence of mass, we obtain for the ratio:

$$d(V_m - V_o) / dV_o \approx (2\pi r_s) r \, dr / 4\pi r^2 dr \approx r_s/2r \approx GM/c^2 r \qquad (9)$$

Thus, we see that the ratio of the differential excess volume in the presence of a mass to the differential volume in the absence of a mass is simply equal in this approximation to the magnitude of the gravitational potential associated with the mass, divided by $c^2$. Thus, this differential volume ratio would correspond to the magnitude of the gravitational potential energy per unit test mass at that particular radial location. So we see that the classical gravitational potential in the space surrounding a mass seems to be interpretable and even visualizable in an explicit and intuitive manner in terms of the local excess spatial volume formed by the distortion of space due to the presence of the mass. Differential excess volume can thus be regarded as a direct geometrical manifestation of gravitational potential.

And how large is this effect? It can be seen to be small, as the ratio in Eqn. (9) is equal to the ratio of half the Schwarzschild radius characterizing the mass to the radius at which the excess volume is determined. We can use values of the physical constants and other relevant quantities from the Appendix to evaluate the fractional excess volume for cases of interest. If we evaluate Eqn. (9) by inserting numerical values appropriate to the surface of the Earth (that is, enter the mass and radius of the Earth and the values for the physical constants G and c), we find for the fractional excess volume 6.9 x $10^{-9}$. Thus, the fractional excess spatial volume created at the surface of the Earth due to the mass of the Earth is about 7 parts in a billion. This rather small value tells us that the behavior of space near the Earth's surface is close to Euclidean. But while this fractional excess volume of 7 billionths of a cubic meter per cubic meter is small, it is not negligible, as it essentially amounts to the local geometrical manifestation representing the gravitational potential here at the Earth's surface.

**Excess spatial volume associated with an isolated mass**

We can continue the investigation of excess volume effects by integrating the differential volume excess in Eqn. (8) so as to obtain an estimate of the total excess spatial volume created by the presence of the mass. While an exact solution could be sought with some manipulation of variables, we will continue using the approximation introduced previously. We find for the excess spatial volume associated with the presence of mass, by integrating Eqn. (8):

$$\delta V = (V_m - V_o) = \int d(V_m - V_o) \approx \int (2\pi r_s) r \, dr \approx \pi r_s r^2 \quad (10)$$

This integral, which we can refer to as the effective excess spatial volume, can give us an estimate of the total difference in volume between the distorted space surrounding a mass and ordinary Euclidean three-dimensional space. The integral requires evaluation between upper and lower radial limits. In regard to the lower limit, it must exceed the very small Schwarzschild radius, and in the absence of information about the mass distribution, the lower limit of integration must also be restricted to the exterior of the mass distribution, to where the Schwarzschild metric is valid. Taking the upper limit of the integral out to an infinite radius would lead to divergence of the integral, and thus to an infinite value for the estimated excess volume. But we do not appear to live in an infinitely extended universe, so a finite upper limit of integration comparable to the size of the universe would seem to be physically justified. Accordingly we will take the upper limit to be the radial extent of the visible universe, which we will designate as $R_U$. In addition, since the lower limit of the integral is very, very small compared to the upper limit, we will neglect the lower limit of the integral. Thus, we find as an estimate for the total excess spatial volume created by an object of mass M:

$$\delta V \approx \pi r_s R_U^2 \approx 2\pi GM R_U^2/c^2 \quad (11)$$

From Eqn. (11), we can see that the excess effective spatial volume associated with a mass will depend linearly on the magnitude of the mass in this approximation. And from Eqn. (11), we also find that the excess spatial volume per unit mass is given at least approximately by a ratio of quantities that are either fundamental physical constants or else are approximately constant at the present epoch:

$$\delta V/M \approx 2\pi G R_U^2/c^2 \quad (12)$$

So, how large an effect is this? We can evaluate it, using values from the tabulation of physical constants and other relevant quantities in the Appendix. If we use a value for the radius of the universe of 1.30 x 10$^{26}$ meters (a commonly used value that corresponds to a 13.75 billion year age of the universe), we can evaluate the right hand side of Eqn. (12) as equal to 7.87x 10$^{25}$ m$^3$/kg. Thus, a kilogram mass would be expected to generate an excess spatial volume of just under 10$^{26}$ cubic meters (which, for orientation purposes, would correspond roughly to about 1/7 of the spatial volume occupied by the sun). For the case of a star of solar mass (about 2 x 10$^{30}$ kg) the volume created would correspond to an excess mass-created volume of about 1.6 x 10$^{56}$ cubic meters (corresponding approximately to the volume of a region of space roughly 500 light years on a side).

We saw earlier that local contributions of excess volume originating from the presence of mass may be small; of the order of some parts per billion at the surface of the earth. However, to examine the total excess volume, we have integrated these small local excess volume contributions throughout very large regions of space, and have found in some cases rather large excess volumes. It now can be seen that this full mass-dependent excess volume effect that seems to be present from general relativity is not a small effect; rather, it is a large effect, and may need to be understood and interpreted on a deeper level.

While these results, based as they are on general relativity, are intended to apply to large macroscopic objects, it may also be of interest to look at what an extrapolation to much smaller masses would suggest. For an electron of mass 9.11 x 10$^{-31}$ kg, we would find an associated effective excess spatial volume $\delta V_e$ = 7.17 x 10$^{-5}$ cubic meter, or about 70 cc; while for a proton of mass 1.67 x 10$^{-27}$ kg, we would find an associated excess spatial volume $\delta V_p$ = 0.13 cubic meter.

We could express these quantities in terms of the fractional spatial volume in relation to the volume of the observable universe. If we evaluate the volume of the universe in terms of a sphere of radius 1.3 x 10$^{26}$ meters in Euclidean space and thus having volume 9.2 x 10$^{78}$ cubic meters, then, a mass of one kilogram would exhibit a fractional spatial volume in relation to the entire volume of the universe of about 9 parts in 10$^{54}$. And for an electron of mass 9.11 x 10$^{-31}$ kg, we would find a fractional volume of about 8 parts in 10$^{84}$. For a proton of mass 1.67 x 10$^{-27}$ kg, we would find a fractional volume of roughly 1 part in 10$^{80}$, approximately the same as for a neutron or an atom of hydrogen. Curiously enough, this extremely large number, 10$^{80}$, also corresponds to an estimate of the number of nucleons contained within our observable universe.[12,13] Thus, this phenomenon of mass-induced excess volume provides another example of large number coincidences connecting cosmological scale effects with atomic scale effects.[12]

**Estimating effects of the presence of multiple masses or an assembly of masses**

The Schwarzschild solution is an exact solution of the equations of general relativity for a particular case: a universe in which only a single isolated mass is present. We will attempt to get a start on at least roughly understanding the creation of additional volume for the case of multiple isolated masses by considering the effects of the masses separately and, to provide a first estimate, combining the results.

Let us designate the individual masses as $M_i$, with values of i ranging from 1 to n, with n being the total number of masses in the ensemble. We will designate the excess volume contributions similarly. Then we can write for a rough initial estimate of the total excess volume, using Eqn. (11) to evaluate the individual contributions:

$$\delta V_{total} = \Sigma\, \delta V_i \approx \Sigma\, 2\pi G M_i R_U^2/c^2 \approx 2\pi G R_U^2/c^2\, \Sigma\, M_i \approx 2\pi G M_{total} R_U^2/c^2 \quad (13)$$

where the total mass $M_{total}$ is equal to the sum of the individual masses, that is, $M_{total} = \Sigma\, M_i$.

We see that, since Eqn. (13) is linear with respect to the mass, to a first approximation the excess space created by all of the individual isolated masses would simply be roughly equal to the excess space that would be created by a single object whose mass is equal to the sum of all of the individual masses.

**Estimating effects of mass throughout the universe**

Let us next consider an assembly of masses that consists of all of the masses that constitute our universe. Using the results of the previous section as the basis for a very rough approximation, we will treat the sum of masses as a sum over all of the masses present in the universe, so that $M_{total} = M_U$. Then, using Eqn. (13), we can estimate the ratio of the excess volume associated with the mass of the universe to a measure of the volume of the universe $V_U$ expressed as $(4\pi/3)R_U^3$. This gives us a rough estimate of the ratio of the excess volume produced by all of the mass in the universe to the volume of the universe:

$$\delta V_{total} / V_U \approx (2\pi G M_U R_U^2/c^2)/(4\pi/3)R_U^3 \approx (3/2)(G/c^2)(M_U/R_U) \quad (14)$$

Since the right hand side of Eqn. (14) consists of fundamental constants and other quantities of known value, we will evaluate it numerically.

The quantity $c^2/G$ is a ratio of physical constants which has the value $1.35 \times 10^{27}$ kg/m. The mass and radius of the universe are less well defined and their values less accurately known; values from the Appendix can be used in their evaluation. If we use for $M_U$ a value of $1.80 \times 10^{53}$ kg (including both stellar masses and dark matter) and for $R_U$ a value of $1.30 \times 10^{26}$ m (corresponding to the age of the universe of 13.75 billion years), then we find for the ratio $M_U/R_U$ a value of $1.38 \times 10^{27}$ kg/m. Thus we find that $\delta V_{total}/V_U \approx (3/2)(1.35 \times 10^{27}$ kg/m$)(1.38 \times 10^{27}$ kg/m$)^{-1} \approx 1.5$. Since we are working at best to orders of magnitude, in the context of the very large numbers that appear in the ratio, the last result can be considered essentially equivalent to 1.

$$\delta V_{total} / V_U \approx 1 \quad (15)$$

So we see that the large numbers that we are working with, almost exactly buck each other out, so that our very crude estimate of the ratio of the excess volume associated with the masses within our universe to the observed volume of the visible universe, is comparable to unity. Thus, our estimate is that the sum of the excess volumes contributed by all of the individual masses in the universe is roughly equal to the total volume of the visible universe itself.

We will examine some possible interpretations and implications of this result in the following section.

**Discussion**

In this section, we will examine some possible implications of these results as well as some of the limitations associated with the assumptions and derivations and interpretation of the results.

A variety of criticisms could be leveled at the rough analysis presented here. It is very crude, and, working in the context of extremely large numbers, we are only considering orders of magnitude at best. However, the results seem interesting, so we will discuss them, even though the present results are clearly quite tentative, and much more careful analysis and interpretation would be desirable.

Here we examine how mass may create excess three dimensional spatial volume in four dimensional space time. However, it should be noted that that excess four dimensional space-time volume is apparently not formed by mass; rather, the space-time volume element is flat, so that there would be no excess four-dimensional volume in space-time.[7]

Consideration of three dimensional volume in four dimensional space time is often downplayed because what is identified as three dimensional spatial volume can be dependent on the frame of reference. Thus, expressing physics in terms of spatial integrals is not covariant because different observers will in general not agree on what they consider spatial. While it seems generally preferable to work with invariants and properties independent of the choice of reference frame whenever that is possible, not all issues can be so addressed; and some problems present preferred reference frames to begin with. Furthermore, we perceive reality in a non-covariant way and develop our intuition based on our perceptions, hence examination of non-covariant features such as the characteristics of three dimensional volume in space-time may be of some value.

Our initial attention has been directed toward examining certain characteristics of the case of an isolated mass, as described by the Schwarzschild solution to the field equations of general relativity. The basic result that we find is that excess space is a concomitant of mass.

Furthermore, we find that small excesses of volume are present throughout space in the vicinity of a mass, and the magnitude of the excess volume corresponds to the local gravitational potential in the case of an isolated mass described by the Schwarzschild solution to the field equations of general relativity. Thus, the concept of local excess volume would appear to be useful pedagogically, in that it presents both a conceptual and a visual interpretation of gravitational potential, and thus would seem to be of possible assistance in developing a more intuitive grasp of this aspect of general relativity and gravitation.

We have gone on to examine potential implications for the universe at large. In addressing and discussing the characteristics of the universe at large, it is more customary and appropriate to address the problem rather more fully, as by looking at a solution to the field equations that takes into account matter density, average curvature, and the cosmological constant or intrinsic energy of the vacuum.[6,14,15] However, sometimes examination of a special case can be informative, even for the larger picture.

The curvature of the observable universe, or the local geometry, seems to be reasonably close to zero, so that space is well described by 3-dimensional flat Euclidean geometry.[15] However flat the averaged out universe is, locally in the vicinity of stars or other massive objects there will be spatial curvature, as we have discussed in connection with the Schwarzschild solution to the gravitational field equations. Furthermore, each such object creates low levels of excess volume throughout the universe. It is the spatial volume associated with these masses that we are seeking to understand in this paper.

In general relativity, this mass-associated spatial volume appears as excess spatial volume in a preexistent space. But if we sum up all of the individual excess volumes associated with all the individual masses in our universe, we find a total volume that is comparable to the volume of the entire observed universe. Is this just a startling coincidence, or what is it telling us?

Would this effect dilute the universe to appreciably lower densities? At present, astrophysics is faced with the dilemma that the overall sum of all of the known normal baryonic matter in the contemporary universe amounts to only about half of what is expected to exist based on observations of atomic matter from the early universe together with computer simulations.[2,3,4] Might the dilution by excess volume that we have been examining here perhaps have relevance to the missing baryonic mass issue in astronomy?

Or is this result telling us something further and different? Could it be that these individual contributions of volume from individual masses could be combining to form the actual full volume of the universe? Could it be that mass is in some sense the ultimate origin of all space? That is, that there might in some sense be no necessity for a preexistent background space, that the background space that we regard as present is a secondary result of a combination of individual contributions from the presence of mass? This of course is a speculation that takes us beyond the usual interpretation of general relativity and beyond our ordinary concept of space.

Another issue originates from the rather remarkable fact that the ratio of mass-induced volume to total volume of the universe approximates unity. That is the observation that this remarkable ratio of near unity would appear to be unique to the present time, due to the fact that it occurs for the present value of the radius of the universe. Thus these results provide another example or aspect of what would appear to be a strange coincidence of physical circumstances having unique relationships at the present time in the history of the universe. At any other time during the history of the universe, the radius of the universe would be different, and would not so closely cancel out the other factors on the right hand side of Eqn. (14), unless one or more of the other parameters ($c$, $G$, $M_U$) also varied in such a manner as to counteract it. So, is there something special and unique about the present era in the history of the universe? Or are some of the quantities that we generally consider as substantially constant, not so? Or what?

We note that somewhat similar concerns have arisen with respect to other issues involving cosmological considerations, including the seeming uniqueness of the present era and certain extremely large number ratios. Some of these issues were raised as long ago as the 1930s by Dirac and subsequently extensively discussed further by Eddington and many others, and additional issues have arisen more recently.[12] It may be that some of the fixes proposed to try to resolve those other issues that involve a seemingly unique role for the present era, and the appearance of very large number ratios, may also be applicable here.

**Summary and conclusions**

As a consequence of general relativity, the presence of mass distorts the space around it, and can affect the volume of three dimensional space around an object. This warping of space by the presence of mass can lead to the presence of additional spatial volume over and above the volume of ordinary three dimensional space in the vicinity of the mass. In the case of an isolated spherically symmetric mass, the ratio of the differential excess volume to the differential volume of flat space appears to provide both a measure of the gravitational potential and an intuitive basis for visualizing the gravitational potential.

Every individual object can be expected to contribute an increment of additional spatial volume to the universe. If we combine the calculated contributions of excess spatial volume produced by all individual masses present in the universe, we find an estimate of the total mass-associated volume that is comparable to the volume of all of space in the observable universe. This suggests the possibility that the dilution of space by excess volume might contribute to explaining the missing mass issue in astrophysics. This also suggests the speculative possibility that space itself might be generated by the masses present.

**Appendix: Physical constants and derived quantities and data for calculations**

While the fundamental physical constants are very accurately known, some quantities used in conjunction with the description of the universe as a whole are characterized by widely different values in the literature, largely because of the use of somewhat differing definitions.[16] Since the calculations in the present paper involve very large (or very small) numbers and are rather crude estimates, we will not sweat an order of magnitude here or there on estimated values.

Here are some values of fundamental physical constants and derived quantities and other data and relevant information.[15,16,17,18]

Gravitational constant: $G = 6.67 \times 10^{-11}$ N m$^2$/kg$^2$; speed of light $c = 3 \times 10^8$ m/s.

Mass of sun: $1.99 \times 10^{30}$ kg; Mass of earth $5.98 \times 10^{24}$ kg; Radius of Earth $6.38 \times 10^6$ m.

Age of the universe: 13.75 billion years or $4.336 \times 10^{17}$ seconds.[18]

Mass of the universe: Some care is required in defining what is meant by the total mass of the observable universe. An estimate based on the total mass for all the stars in the observable universe is $3 \times 10^{52}$ kg.[16] Dark matter is reportedly present at about 5 times the incidence of hadronic matter in the universe.[19] If we include dark matter in addition to stellar matter in estimating the mass of the universe, that would lead to an estimate of approximately $1.8 \times 10^{53}$ kg for the mass of the universe.

Radius of the universe: Widely varying values can be found in use as a measure of the radius of the universe.[16] The age of the universe is 13.75 billion years, and this is often used as an estimate of the radius of the observed universe.[16] Estimating the radius of the universe as the product of its age equivalent to 13.75 billion years and the speed of light, a radius of $1.30 \times 10^{26}$ meters is obtained. Another estimate is provided by the comoving distance from Earth to the edge of the observable universe, which is about $4.35 \times 10^{26}$ meters in any direction.[16]

Volume of the universe: The age of the universe is 13.75 billion years, and this is often used as an estimate of the radius of the universe as $1.3 \times 10^{26}$ meters; assuming a Euclidean space, this would lead to a volume of about $9.2 \times 10^{78}$ cubic meters. The geometry of the universe can also be represented in a system of comoving coordinates in which the expansion of the universe can be ignored; these comoving coordinates form a single frame of reference in which the universe has a static geometry of three spatial dimensions.[15] The comoving distance from Earth to the edge of the observable universe is about $4.35 \times 10^{26}$ meters in any direction. The observable universe would thus be a sphere with a radius of $4.35 \times 10^{26}$ meters with respect to commoving coordinates.[16]

volumemassGRnu1.doc
13 May 2011 draft